\documentclass[twocolumn, switch]{article} 
\usepackage{preprint}

\usepackage[utf8]{inputenc} 
\usepackage[T1]{fontenc}    
\usepackage{amsmath}
\usepackage{amsfonts}
\usepackage{mathpazo}
\usepackage{inconsolata}                             
\usepackage{xcolor}
\usepackage[hyphens,obeyspaces,spaces]{url}
\usepackage{hyperref}
\usepackage{multicol}
\usepackage{xspace}                                
\usepackage[detect-all]{siunitx}
\usepackage[font=small]{caption}
\usepackage{bm}
\usepackage{microtype}      
\usepackage{float}			
\usepackage{authblk}

\graphicspath{{figures/}}                           
\newcommand{\subrm}[1]{\ensuremath{_{\text{#1}}}}
\newcommand{\transpose}[1]{#1^{\scriptscriptstyle\mathsf{T}}}
\providecommand{\degree}{\ensuremath{^\circ}\xspace}    
\renewcommand{\degree}{\ensuremath{^\circ}\xspace}      
\newcommand{\figref}[1]{Fig.~\ref{#1}}

\newcommand{\vv}[1]{\bm{\mathrm{#1}}}                   
\newcommand{\adjoint}[1]{#1^{\raisebox{\depth}{\rotatebox{180}{\small\sf A}}}}

\def\citenum#1{\def\@cite##1##2{##1}\cite{#1}}

\makeatletter      
   \newcommand{\figcaption}{\captionsetup{type=figure} \def\@captype{figure}\caption}            
   \newcommand{\tabcaption}{\captionsetup{type=table} \def\@captype{table}\caption}            
   \renewcommand{\@biblabel}[1]{#1.}    
\makeatother

\title{Wave description of geometric phase}

\author[1]{Luis Garza-Soto}
\author[1\thanks{\tt{nh@hagenlab.org}}]{Nathan Hagen}
\author[2]{Dorilian Lopez-Mago}
\author[1]{Yukitoshi Otani}

\affil[1]{Department of Optical Engineering, Utsunomiya University, 7-1-2 Yoto, Utsunomiya, Tochigi 321-8585 Japan}
\affil[2]{Tecnologico de Monterrey, Escuela de Ingenier{\'\i}a y Ciencias Ave.\ Eugenio Garza Sada 2501, Monterrey, N.L., M{\'e}xico, 64849}

\begin{document}

\twocolumn[ 
  \begin{@twocolumnfalse} 
  
\maketitle

\begin{abstract}
   Since Pancharatnam's 1956 discovery of optical geometric phase, and Berry's 1984 discovery of geometric phase in quantum systems, researchers analyzing geometric phase have focused almost exclusively on algebraic approaches using the Jones calculus, or on spherical trigonometry approaches using the Poincar{\'e} sphere. The abstracted mathematics of the former, and the abstracted geometry of the latter, obscure the physical mechanism that generates geometric phase. We show that optical geometric phase derives entirely from the superposition of waves and the resulting shift in the location of the wave maximum. This wave-based model provides a way to visualize how geometric phase arises from relationships between waves, and from the transformations induced by optical elements. We also derive the relationship between the geometric phase of a wave by itself and the phase exhibited by an interferogram, and provide the conditions under which the two match one another.
\end{abstract}
\vspace{0.35cm}

  \end{@twocolumnfalse} 
] 

\section{Introduction}

As we have come to learn more about geometric phase over the past 38 years,\cite{Pancharatnam1956,Berry1984} we have acquired the ability to calculate it for an increasing array of circumstances. Yet, even now we lack a physical model for visualizing geometric phase. For polarization optics, the Poincar{\'e} sphere has been a widely used tool to visualize the various calculations and to describe how the geometric phase relates to transformations of polarization states, but its use opens up even more questions.\cite{Courtial1999} For example, when using the Poincar{\'e} sphere to calculate the geometric phase, the typical definition states that when a state of polarization (SOP) undergoes a series of transformations, the various transformations induce a phase delay $\gamma$ to the wave, equal to half the solid angle subtended on the surface of the sphere: $\gamma = \Omega / 2$. However, when drawing the curves used to delineate the spherical area, researchers often ignore the fact that the actual physical path taken by the polarization state is not that of the curves they draw. Thus, while the calculation proceeds correctly, the underlying physics that drives the calculation methods remains unclear. 

The following presents a list of some of the curious rules under which the Poincar{\'e} sphere areas must be calculated in order to obtain correct estimates of the geometric phase:
\begin{enumerate}
   \item Pancharatnam's original work never considered a cycle of states, but rather considered only two input states $A$ and $B$, and the state generated by their sum, $C = A + B$.\cite{Pancharatnam1956} The states $A$ and $B$, together with $C'$ (the antipode to $C$), form a triangle of points on the Poincar{\'e} sphere from which one can calculate the subtended solid angle, $\Omega$.\cite{Aravind1992}
   \item The subtended angle formed by the actual physical path of the polarization state is generally not the correct solid angle $\Omega$ needed to get the correct geometric phase. Rather, for the correct $\Omega$ one must use the shortest geodesic arc connecting the pair of SOPs before and after each homogeneous polarization element.\cite{Courtial1999,Zhou2020} In addition, if the physical path of the polarization state \textit{is} a geodesic, such as when using a half-wave plate (HWP), then one must use the geodesic that coincides with the physical path.
   \item One of the consequences of separating the physical path of the polarization state from the geodesic path of calculation is that it becomes difficult to determine the value of the geometric phase inside optical elements such as a linear retarder. Rather, analyses are generally limited to calculating $\gamma$ before entering or after leaving an element, but not the continuous changes that occur while propagating through it.
   \item The areas on the Poincar{\'e} sphere are actually signed areas --- negative if clockwise, positive if anticlockwise, and that if a path crosses itself then one can have positive and negative areas partially cancelling one another.\cite{Kurzynowski2011,Martinez-Fuentes2012,Gutierrez-Vega2020}
   \item While the spherical angle is usually defined using a closed cycle of polarization states,\cite{Arteaga2020,Jisha2021} a cycle of states can never in practice be exactly closed, it is necessary to have a procedure for calculating the geometric phase that allows for a set of states that are only approximately closed, or not closed at all.\cite{Dijk2010a} This procedure is as follows: the phase of non-closed loop of states $A \to B \to C$ is equal to the Pancharatnam–Berry phase related to the closed loop $A \to B \to C \to A$, plus the phase of the Pancharatnam connection related to the projection of $A \to C$.\cite{Martinez-Fuentes2012}
   \item Under some conditions, the geometric phase can undergo a $\pi$ shift singularity.\cite{Bhandari1991,Galvez2002,Dijk2010a,Zhou2020}
\end{enumerate}
While the above rules are useful guides to obtaining the correct result, the abstraction of the spherical geometry leaves little insight into the physical origin guiding why these rules must exist, and can even hinder understanding. Definitions found in the existing literature, such as ``Geometric phase is a consequence of parallel transport in a curved topology''~\cite{Galvez1999a} or similar abstract treatments~\cite{Cisowski2022} are not exactly wrong, but they can be considered misleading, in that much simpler processes are actually at work. 
In the discussion below, we show that simply analyzing the locations of wave maxima, without the abstraction of the Poincar{\'e} sphere or of matrix algebra, generates the geometric phase properties while also allowing one to visualize the underlying physics. Moreover, in contrast to existing methods, our wave-model of the geometric phase provides a clear means of defining the geometric phase at any point in its propagation through an optical system --- not only at points before and after traversing a homogeneous optical element, but inside the element as well.

Section~\ref{sec:1D} starts with the simplest possible case --- the superposition of waves in 1D --- and introduces a reference plane from which we can define geometric phase. When two waves are superposed, the location of the wave amplitude peak of their sum depends on both the relative phases of the input waves and on their relative amplitudes. This change in position from the input reference plane to output state peak location is the geometric phase. 

Section~\ref{sec:2D} generalizes this superposition analysis to 2D, and shows how we can define the phase of an arbitrarily polarized wave with respect to the two orthogonally polarized waves that compose it. This section also introduces a visualization aid for locating the polarized wave peaks.


In Sec.~\ref{sec:retarder}, we provide an example of propagating polarized waves through a linear retarder. Although previous authors have claimed that single linear retarders cannot introduce geometric phase,\cite{Kurzynowski2011} more recent work has argued that indeed they can,\cite{Gutierrez-Vega2011} and our work as well shows that such retarders do in general cause a global phase shift.

Finally, we note that the above definitions of geometric phase involve propagating waves, and so these are theoretical constructs that are not directly measurable. Section~\ref{sec:interferogram} introduces a model for interfering a sample wave with a reference wave, generating a stationary interferogram that is measurable. The resulting interferogram includes the geometric phase $\gamma$ previously obtained for propagating waves. However, one finds that if the two arms of the interferometer deliver waves of different polarization states, then the interferogram phase will differ from the wave geometric phase. This is similar to the requirement  sometimes stated in the historical literature that the geometric phase is only defined for ``closed loop'' transformations of the polarization state. However, we show that this condition is in fact too restrictive. Rather than requiring that the two beams share the same polarization state, we find that the two beams need only share the same ellipticity.

\section{Wave composition in 1D}\label{sec:1D}

It is a well-known property of sinusoidal waves that the sum of any two waves is also a sinusoid, with an amplitude and phase that depends on the states of the input waves. For two co-propagating electromagnetic waves $E_1$ and $E_2$ of arbitrary amplitude and phase $E_1 = A_1 \cos (kz - \omega t - \phi_1)$ and $E_2 = A_2 \cos (kz - \omega t - \phi_2)$, their sum is given by the Harmonic Addition theorem as~\cite{Hecht1990,Weisstein2003}
\begin{equation}\label{eq:sum1d}
   E_1 + E_2 = A_3 \cos (kz - \omega t - \gamma) \, ,
\end{equation}
where $k = 2 \pi / \lambda$ is the wavenumber, $\omega$ the angular frequency, and where
\begin{align}
   A_3^2 &= A_1^2 + A_2^2 + 2 A_1 A_2 \cos (\phi_2 - \phi_1) \, , \label{eq:A3} \\
   \tan \gamma &= \frac{A_1 \sin \phi_1 + A_2 \sin \phi_2}{A_1 \cos \phi_1 + A_2 \cos \phi_2} \, , \label{eq:tan_phi3}
\end{align}
give the resultant amplitude $A_3$ and phase $\gamma$. This can also be generalized to the case of adding $N$ arbitrary waves, as shown in Appendix~\ref{app:Nwaves}. Note that by using cosines to represent our two waves, we have implicitly chosen the wave peaks to indicate the phase origin ($\phi = 0$) position. A phase advance is given by $\phi > 0$, and a phase delay by $\phi < 0$.

\begin{figure}[h]
    \centering
    \includegraphics[width=0.9\linewidth]{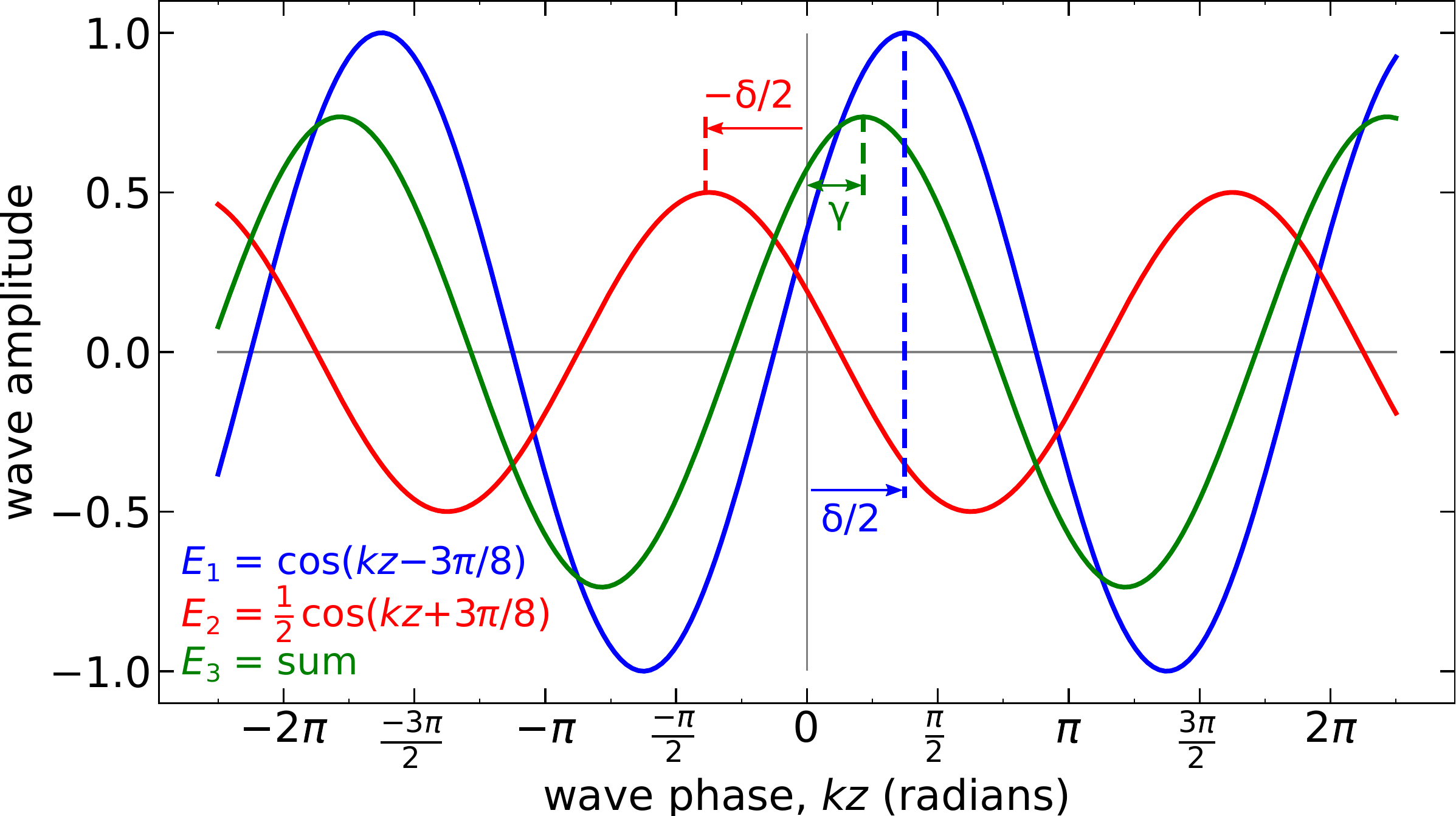}
    \figcaption{Superposition of two waves with amplitudes $A_1 = 1$ (blue) and $A_2 = 0.5$ (red) and a phase difference $\delta = 3 \pi/4$ between them. The peak location $\gamma$ of the resultant wave (green) shifts towards the wave with greater amplitude (blue). The phase of each wave ($\phi_i = \pm 3 \pi/8$) is defined with respect to a reference position ($z = 0$ here) given by the midpoint between the two component wave peaks. The phase $\gamma = 38.8\degree$ of the sum wave's peak is given by \eqref{eq:tan_phi3_symmetric}. (See Visualization~1 for an illustration of wave composition for a range of phase delay values.)}
    \label{fig:adding_1Dwaves}
\end{figure}

If we choose to locate our reference plane halfway between the $E_1$ and $E_2$ peaks, the phase difference $\delta = \phi_2 - \phi_1$ between the two waves is split in half on each side of the reference plane, such that $\phi_1 = \delta/2$ and $\phi_2 = -\delta/2$. In this case, the phase $\gamma$ of the sum wave simplifies to
\begin{equation}\label{eq:tan_phi3_symmetric}
   \tan \gamma = \tan (\delta/2) \frac{A_1 - A_2}{A_1 + A_2} \, .
\end{equation}
If the amplitudes of the two input waves are equal ($A_1 = A_2 = A$), using the product identity for cosines with \eqref{eq:A3} obtains
\begin{equation}\label{eq:E3}
   E_3 = 2A \cos (\delta/2) \cos (kz) \, .
\end{equation}
Thus, when the two input wave amplitudes are equal, the sum wave amplitude becomes $A_3 = 2A \cos (\delta/2)$ and the sum wave phase is exactly at the midpoint between the phases of the input waves ($\gamma = 0$). The value of $\gamma$ indicates the location of the sum wave peak relative to the reference plane. If the two input wave amplitudes are not equal ($A_1 \neq A_2$), then Eq.~(\ref{eq:tan_phi3_symmetric}) indicates that the phase of the sum wave shifts towards the phase of the input wave with greater amplitude. This situation is illustrated in Fig.~\ref{fig:adding_1Dwaves}, where we see that the phase of the green sum wave moves closer to the blue wave phase due to the latter's higher amplitude.

\vspace{\baselineskip}

In \eqref{eq:tan_phi3_symmetric}, one might be concerned about the case when $A_1 = -A_2$, in which case the denominator in the equation becomes zero. However, our choice of the phase delay between the two component waves allows the negative sign to be replaced with an extra phase delay of $\delta \to \delta + \pi$, while maintaining positive amplitudes. This avoids possible division by zero.

If the phase separation $\delta$ between the two input waves is greater than $\pi$, then an ambiguity arises. Since we cannot generally measure the absolute phase of a wave, but can only measure phase differences between waves, when $\delta > \pi$ the nearest $E_2$ peak (red wave) with respect to the $E_1$ (blue wave) peak will no longer be to the left of the origin, but rather to the right. In our convention, we always choose the nearest pair of peaks, so that $\delta$ for $E_3$ (green wave) can never exceed $\pi$. As we will see below, this change from using the left peak to the right peak corresponds exactly to behavior discussed in the existing literature on the geometric phase, in which $\gamma$ can be made to switch instantaneously between $+\pi/2$ and $-\pi/2$.\cite{Dijk2010a}


\section{Wave composition in 2D}\label{sec:2D}

Unlike the 1D case, we cannot simply add orthogonally-vibrating waves and look for the resulting peak. Instead, we form the polarization ellipse traced by the electric vector produced by adding the two orthogonal waves. The point at which the electric vector aligns with the major axis of the ellipse is designated to be the location of the 2D wave peak --- its phase. 
Generalizing our approach to plane waves in 2D involves the same basic procedure as in 1D, but we will see that we must work with the squares of the electric fields rather than the fields themselves. 

To start, we use the midpoint between the two orthogonal component waves' peaks to define a reference plane, and define the geometric phase as the phase difference of the elliptical wave's sum peak from the reference plane. While this is easy to do when working with linear polarization states, waves in 2D are elliptically polarized in general --- their instantaneous electric field vectors have an orientation that varies with time. Some polarization components are more conveniently represented using a basis of two elliptical states, but if we want to define a ``phase'' for representing an elliptically polarized wave, we need a convention. Pancharatnam proposed the natural choice of defining $\phi = 0$ at the position where the electromagnetic vector is at its maximum positive displacement from the axis. This is also known as \textit{Pancharatnam's connection}, and can also be stated as defining the relative phase $\phi$ between two interfering waves such that $\phi = 0$ (the two waves are ``in phase'') when their interference is maximally constructive.\cite{connection} 

For a general elliptical state of polarization propagating along the $z$-axis, the real-valued electric fields of the $E_x$ and $E_y$ components are given by
\begin{align}
   E_x &= A_x \cos (kz - \omega t - \phi_x) \, , \\
   E_y &= A_y \cos (kz - \omega t - \phi_y) \, .
\end{align}
At any given position $z$ and time $t$ the vector magnitude of the sum of these two components is
\begin{equation}\label{eq:absE}
   |E| = \sqrt{A_x^2 \cos^2 (kz - \omega t - \phi_x) + A_y^2 \cos^2 (kz - \omega t - \phi_y)} \, .
\end{equation}
In order to locate the $z$-position at which the electric field magnitude is maximum, we take the derivative with respect to $z$ (at constant $t$) and search for the location where the derivative is zero. After some algebraic work, the derivative becomes
\begin{align}
    \frac{d|E(z)|}{dz} &= -k A_x^2 \sin \big[ 2 (kz - \omega t - \phi_x) \big] \notag \\
      &\qquad - k A_y^2 \sin \big[ 2 (kz - \omega t - \phi_y) \big] \, . \label{eq:dE_dz}
\end{align}
We can see that this has a similar form to \eqref{eq:sum1d}, and so it comes as little surprise that after some algebra we obtain a similar expression for $\gamma$:
\begin{equation}\label{eq:tan2L}
   \tan \gamma = \frac{A_x^2 \sin \phi_x + A_y^2 \sin \phi_y}{A_x^2 \cos \phi_x + A_y^2 \cos \phi_y} \, .
\end{equation}
As we show in App.~\ref{app:compare}, the expression \eqref{eq:tan2L} agrees with existing expressions in the geometric phase literature obtained via Jones calculus.\cite{Lopez-Mago2017}

\begin{figure}[h]
    \centering
    \includegraphics[width=0.55\linewidth]{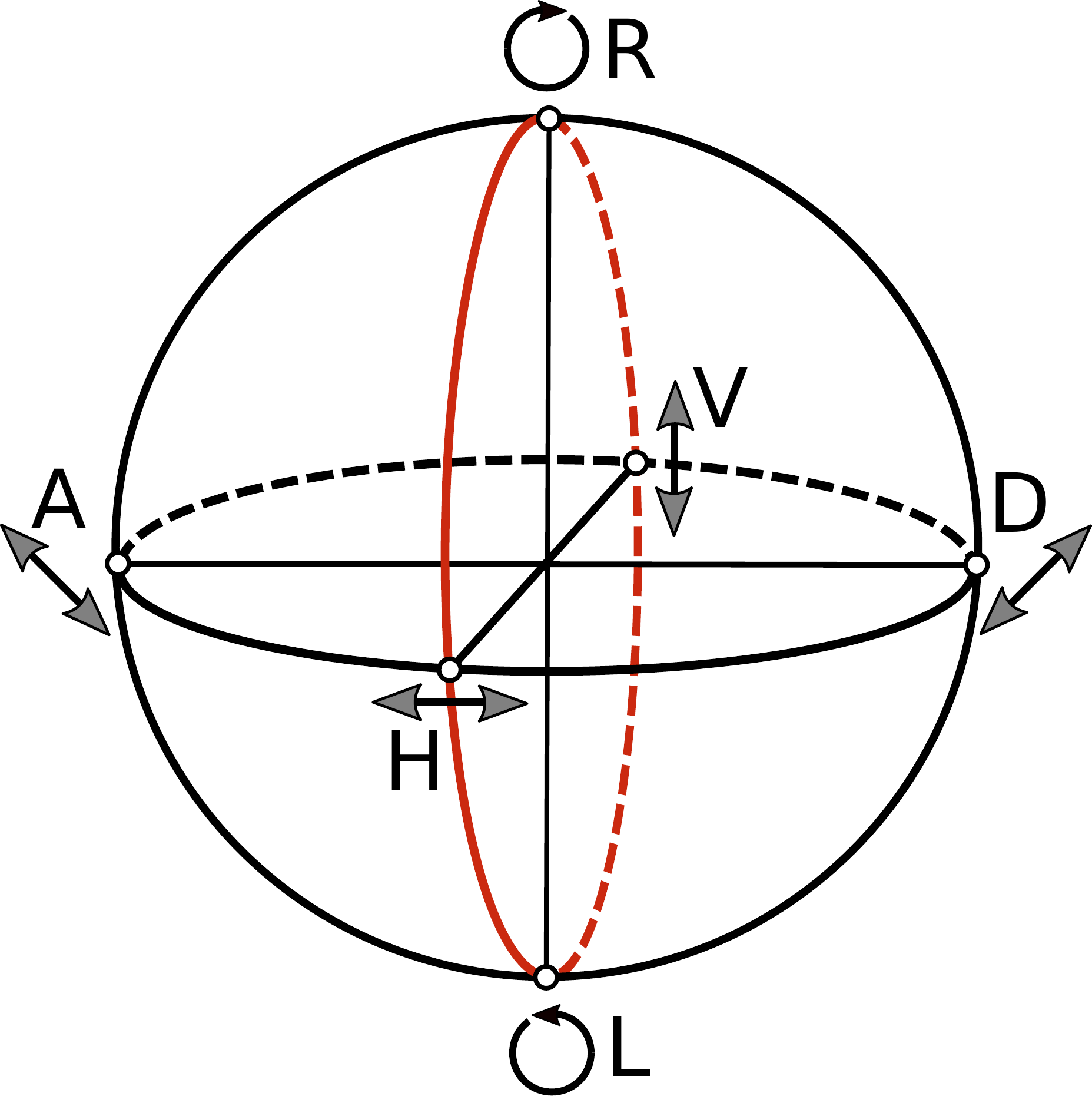}
    \figcaption{The great circle drawn in red on the Poincar{\'e} sphere indicates the states of polarization for which $\gamma = 0$ via \eqref{eq:tan2L}. Dashed curves indicate locations that are on the opposite face of the sphere. The axis labels (H, V, D, A, R, L) indicate the horizontal, vertical, diagonal, antidiagonal, right-circular, and left-circular polarization states.}
    \label{fig:poincare_inphase}
\end{figure}

\begin{figure*}[h]
   \centering
   \includegraphics[width=0.85\linewidth]{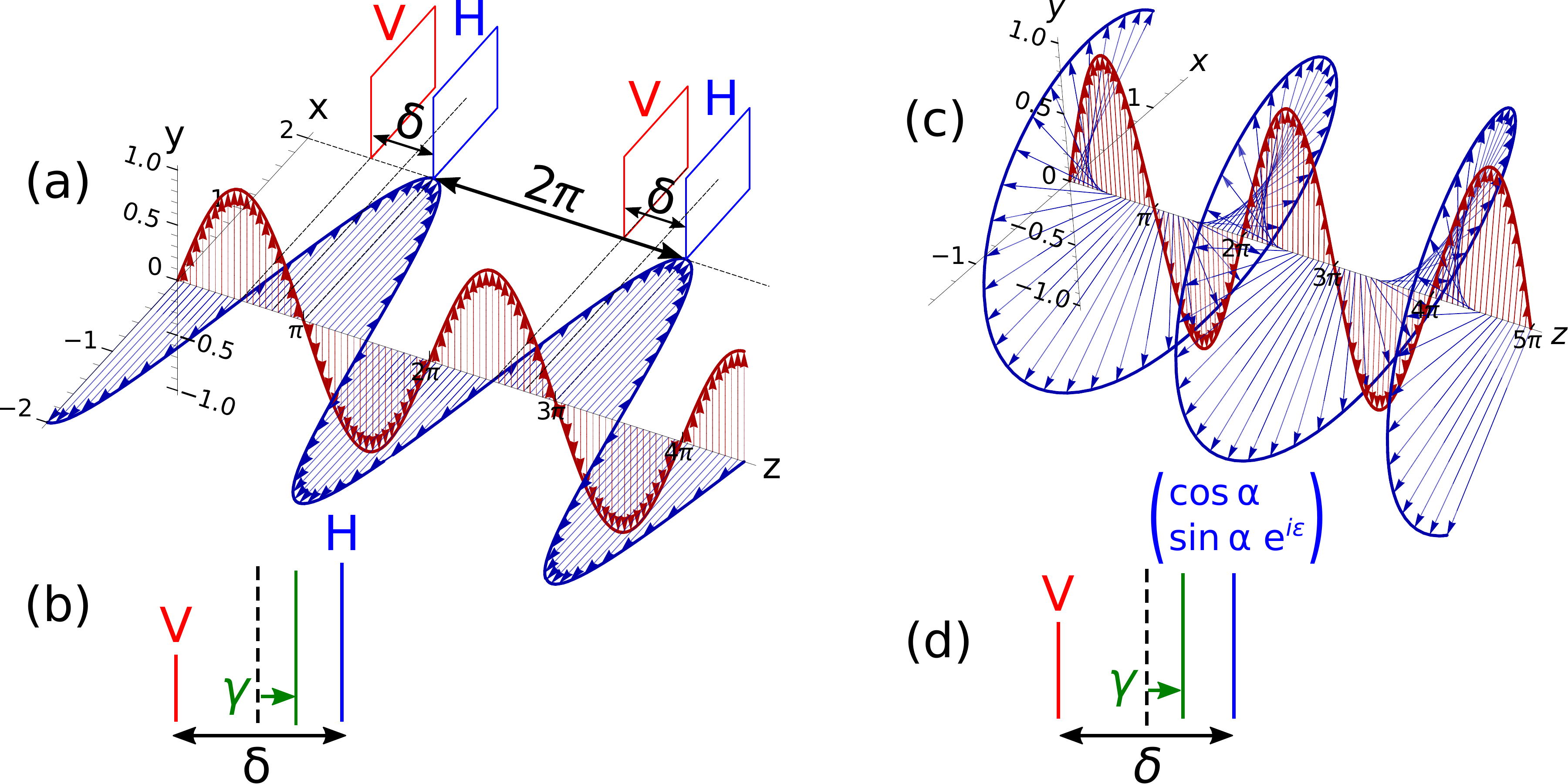}
   \caption{(a) Two waves oscillating in the $x$ and $y$ axes, with $A_x = 2$, $A_y = 1$, and $\phi_x = \pi$, $\phi_y = \pi/2$. The H and V ``wavefronts'' are drawn at the wave peaks. (b) The wavefronts from (a). (c) A vertically polarized and an elliptically polarized wave, with amplitudes $A_1 = 1$, $A_2 = 1.5$ and phases $\phi_1 = \pi/2$, $\phi_2 = 5 \pi/8$. (d) The wavefronts from (c).}
   \label{fig:adding_2Dwaves}
\end{figure*}

As with wave superposition in 1D, if there is no fixed reference plane, then it is convenient to choose the reference plane to be such that $\phi_x$ and $\phi_y$ are symmetrically displaced with the reference halfway between them: $\phi_x = -\phi_y = \delta / 2$. In such a case, \eqref{eq:tan2L} simplifies to
\begin{equation}\label{eq:tan2L_symmetric}
   \tan \gamma 
      = \tan (\delta/2) \, \frac{A_x^2 - A_y^2}{A_x^2 + A_y^2} \, .
\end{equation} 
Comparing the 2D Pancharatnam phase \eqref{eq:tan2L_symmetric} with the 1D Pancharatnam phase \eqref{eq:tan_phi3_symmetric}, we can see that they have almost the same form, but with the 1D expression's amplitudes $A_1,A_2$ replaced with the 2D expression's $A_x^2, A_y^2$. 

Figure~\ref{fig:adding_2Dwaves}(a) illustrates two co-propagating linearly-polarized waves, with electric field amplitudes $A_x = 2$ and $A_y = 1$. At the location where each wave reaches its maximum, we draw a plane to indicate the wave position --- the ``wavefront''. For phase difference $\delta \neq 0$, the superposition of these two waves will be an elliptically polarized wave (not shown in the figure) that has its maximum displacement at $\gamma$ with respect to the midpoint reference plane (the dashed line). The location of the sum-wave's peak is indicated by $\gamma$.

Figure~\ref{fig:adding_2Dwaves}(b) shows a simplified drawing where the wavefronts are represented by vertical lines. The length of each line indicates the amplitude of the wave, and the horizontal distance between the two lines represents the phase difference $\delta$ between the two components. The reference plane is indicated by a vertical dashed line, given by the midpoint between the peaks of the two component waves.

Figure~\ref{fig:adding_2Dwaves}(c) shows the case of superposing a vertically polarized and an elliptically polarized wave. Once again, the wave positions are defined with respect to the field vector maxima. (For the elliptical wave, this is maximum is equal to half the length of the polarization ellipse's major axis.) The corresponding wavefront representation is shown in Figure~\ref{fig:adding_2Dwaves}(d).

\begin{figure}[h]
    \centering
    \includegraphics[width=0.85\linewidth]{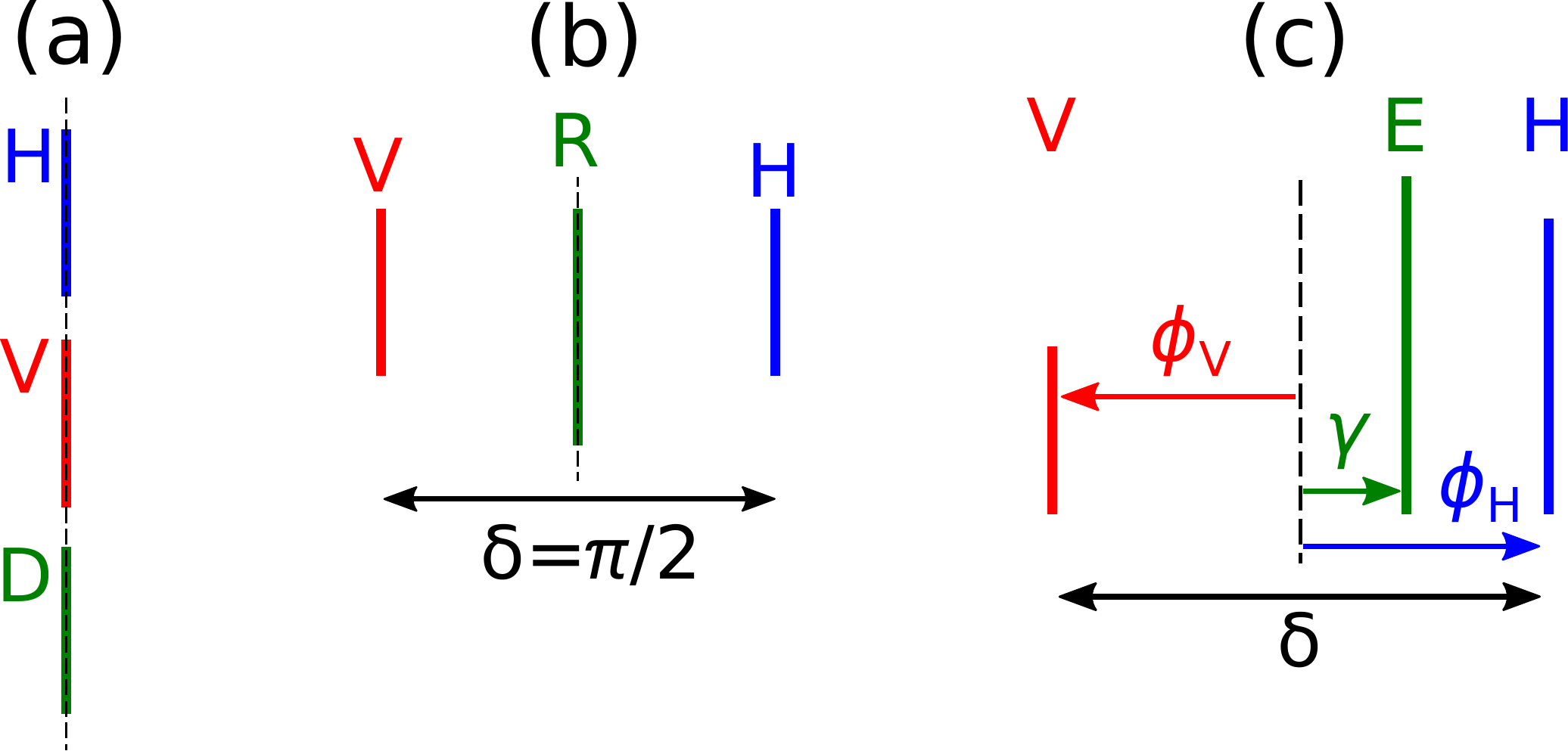}
    \figcaption{Using the wavefront representation to aid wave composition calculations. (a) Input horizontally-polarized (H) and vertically-polarized (V) waves sum to create a diagonally-polarized (D) wave oriented at $45\degree$, in phase with the two input waves. (b) The same input waves as in (a) but now with a phase difference $\delta = \pi/2$ between them. The resultant sum wave is elliptically polarized in general, but right-circularly polarized (R) if $A\subrm{H} = A\subrm{V}$. Since the two input waves are in phase with one another, the sum wave is in phase with them. (c) Input waves of arbitrary amplitudes $A\subrm{H}$ and $A\subrm{V}$ and phase difference $\delta$. The resultant elliptical sum wave (E) has position $\gamma$ relative to the midpoint between the wavefront positions of the two input waves.}
    \label{fig:wavefronts1}
\end{figure}

Figure~\ref{fig:wavefronts1} shows three examples how this wavefront representation can be used to help visualize wave composition and the resulting geometric phase $\gamma$. Figure~\ref{fig:wavefronts1}(a) shows a horizontal polarized wave and a vertical polarized wave that are in phase and have the same amplitude. Their sum produces diagonally polarized light (azimuth angle 45\degree) with respect to the $x$ axis. From \eqref{eq:tan2L}, we find that the position of the wavefront of this sum wave is given by $\gamma = 0$ since the relative phase between the components is $\delta = 0$. This means that the diagonally-polarized sum wave is in phase with both of the input component waves.

Figure~\ref{fig:wavefronts1}(b) shows a horizontally polarized wave and a vertical polarized wave with equal amplitudes, and a phase difference $\delta = \pi/2$. The addition of the two waves produces right-circularly polarized light. This time \eqref{eq:tan2L} gives $\gamma = 0$ because the amplitudes of the two component waves are equal. Once again, the sum wave is in phase with the inputs.

Figure~\ref{fig:wavefronts1}(c) shows the general case for summing two waves using an $x$-$y$ basis. The two input waves have different amplitudes $A\subrm{H} \neq A\subrm{V}$, and a phase difference $\delta$. The sum wave will have a phase $\gamma$ lying between the two input wavefront positions.

\vspace{\baselineskip}

While \eqref{eq:tan2L} gives the expression for the geometric phase, it does not yet represent a physically measurable quantity, because an interferogram is needed to detect a global phase shift. Section~\ref{sec:interferogram} discusses how this expression can be generalized for the addition of 2D waves to produce a measurable quantity. However, we can note that the expression given in \eqref{eq:tan2L_symmetric} contains much of the behavior of geometric phase that excites curiosity. For example, if the waves were to pass through a retarder $R$, and we want to compare the input wave phase to the output wave phase, we cannot simply subtract their phases because the retardance induced by $R$ will be present inside the tangent function of \eqref{eq:tan2L_symmetric}. This induces a nonlinear behavior in the phase, such that the path by which one reaches the output has an effect on the calculation.

\section{The effect of polarization components on geometric phase}\label{sec:retarder}

In order to demonstrate how our wave-based approach to geometric phase can deal with optical elements, we use our wavefront visualization technique (\figref{fig:wavefronts1}) and analyze the effect of a linear retarder on the wavefront phase.

\begin{figure}
   \centering
   \includegraphics[width=0.95\linewidth]{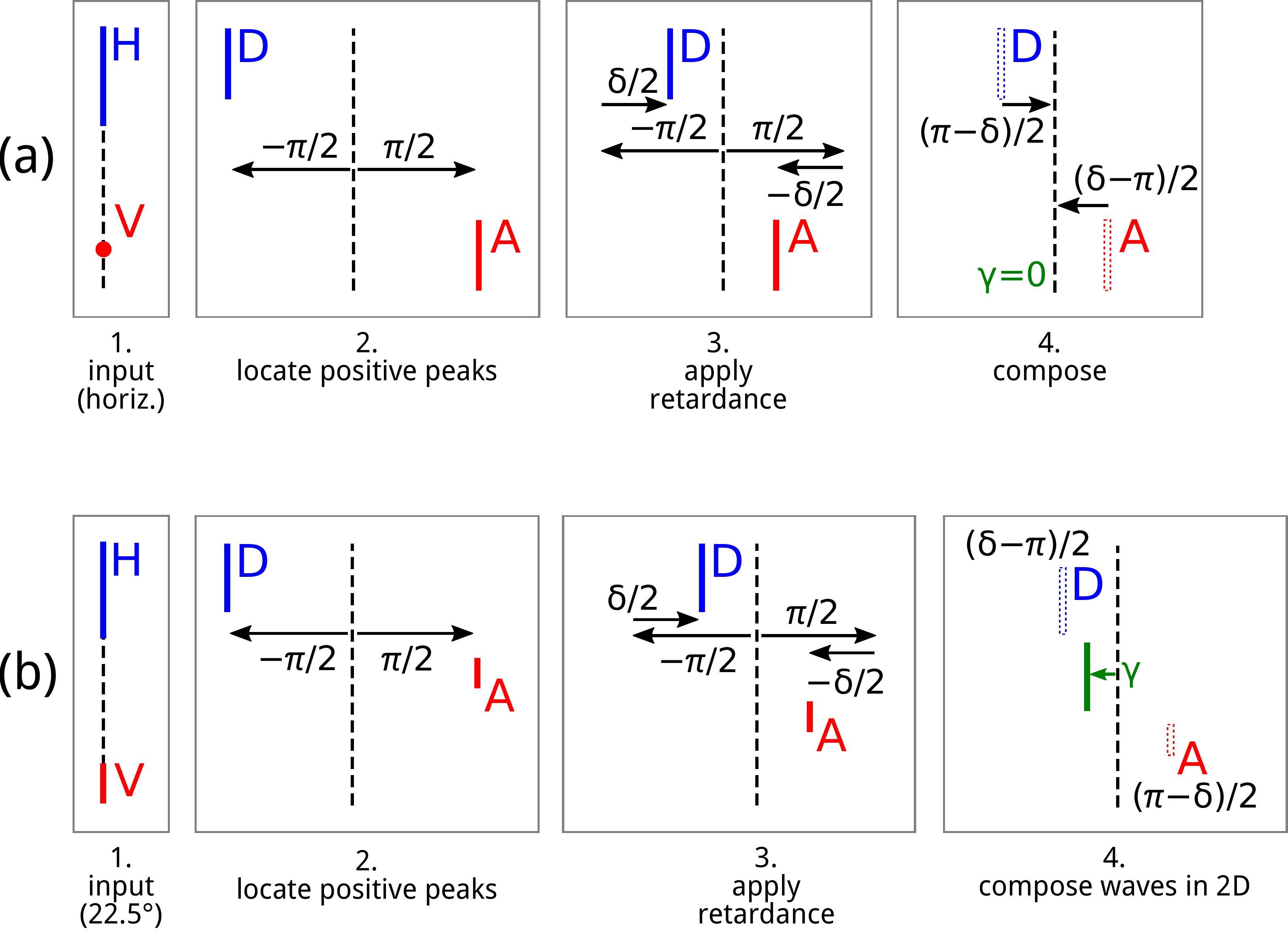}
   \figcaption{Diagram for visualizing the propagation of a polarized wave through a linear retarder (retardance $\delta$) with its fast axis oriented at 45\degree: ($a$) a horizontally-polarized (H) wave input, ($b$) a 22.5\degree linearly-polarized wave input.}
   \label{fig:retarder_phase}
\end{figure}

Figure~\ref{fig:retarder_phase}(a) shows each step for calculating the phase for a horizontally-polarized wave propagating through a linear retarder whose fast axis is oriented at 45\degree:
\begin{description}
   \item[Step 1] The input wave is horizontally polarized, so the V-wave component is zero.
   \item[Step 2] Transform the horizontally-polarized wave into the eigenbasis of the retarder: the diagonal and antidiagonal (D,A) linear polarization states. Since a projection from a linear polarization state onto any orthogonal linear polarization basis produces only 0 phase shift, the D and A waves have to be symmetrically shifted with respect to the reference plane. While it is also possible to use a $-\pi$ phase shift for the A-wave, this creates a distance of more than $\pi$ between positive wave peaks. Our convention is to shift all wavefronts such that phase differences between the positive wave peaks being added never exceeds $\pi$.
   \item[Step 3] Apply the retardance to the two waves, by advancing the D wave and delaying the A wave symmetrically. (The propagation phase due to the retarder is incorporated into the dynamic phase and, as we will see in Sec.~\ref{sec:interferogram}, does not play a role in the final result.) 
   \item[Step 4] Compose the D and A waves to determine the phase of the resulting elliptical state. Since the peak of the resulting state coincides with the location of the original reference plane, we find $\gamma = 0$.
\end{description}

\vspace{\baselineskip}

In this example the D and A waves superposed after passing through the retarder are of equal amplitude. As a result, the phase of their sum is located at the midpoint between the two input peaks, and therefore $\gamma = 0$. In fact, for a 45\degree orientation linear retarder, this situation of zero geometric phase will occur for any input polarization state that is located on the red circle drawn in \figref{fig:poincare_inphase}.\cite{Lopez-Mago2017}

In order to demonstrate the method by more conventional means, we follow the same five steps above with numerical calculations. Step~1 begins with a polarization state $\vv{E}\subrm{H} = \transpose{(1 \ 0)}$. In order to apply the waveplate retardance, we need to decompose this into the (D,A) basis. Doing so gives two waves, $\vv{E}_1$ and $\vv{E}_2$ given by
\begin{equation}
   \vv{E}_1 = \frac{1}{\sqrt{2}} \vv{E}\subrm{D} \, , \quad \vv{E}_2 = -\frac{1}{\sqrt{2}} \vv{E}\subrm{A} \, ,
\end{equation}
where $\vv{E}\subrm{D}$ and $\vv{E}\subrm{A}$ are normalized Jones vectors for the D and A polarization states. Once we apply the retardance (Step~3), these two waves become
\begin{equation}
   \vv{E}_1 = \frac{1}{\sqrt{2}} e^{-i \delta / 2} \vv{E}\subrm{D} \, , \quad \vv{E}_2 = -\frac{1}{\sqrt{2}} e^{i \delta / 2} \vv{E}\subrm{A} \, .
\end{equation} 
Here we have ignored the dynamic phase --- an assumption that is justified later in \eqref{eq:interferogram}.

In Step~4, we compose the D- and A-wave into a single 2D polarized wave using \eqref{eq:tan2L}, with
\begin{equation}
   \bar{A}^2 = \Big( \frac{1}{\sqrt{2}} \Big)^2 + \Big( \frac{-1}{\sqrt{2}} \Big)^2 = 1
\end{equation} 
and
\begin{equation}\label{eq:gamma-bar}
   \bar{\gamma} 
      = \arctan \Big[ \frac{\sin (\delta / 2) + \sin (-\delta / 2)}{\cos (\delta / 2) + \cos (-\delta / 2)} \Big] = 0 \, ,
\end{equation} 
where the overbars indicate that the amplitude and phase refer to the 2D wave rather than to either of its 1D component waves. (Note that the equal amplitudes of D- and A-waves have allowed the amplitudes to factor out from the expression.)

From \eqref{eq:gamma-bar}, we see that the geometric phase $\bar{\gamma}$ produced by the waveplate in this case is zero --- a result that holds true for any input polarization state that lies on the red circle drawn in \figref{fig:poincare_inphase}. For an input SOP that does not lie on the red circle, the geometric phase will not be zero. This result agrees with the previous findings of Ref.~\citenum{Lopez-Mago2017}.

The step-by-step process outlined above produces the same geometric phase as is obtained from the standard Jones vectors.\cite{Gutierrez-Vega2011} One drawback of the Jones vector approach is that so far it has only been applied to calculate $\gamma$ before and after homogeneous optical elements, and not within them. In this approach, for the situation shown in \figref{fig:retarder_phase}(a), the Jones vectors of the input and output waves are $\vv{E}_a = \transpose{(1 \ \ 0)}$ and
\begin{equation}
   \vv{E}_b = (1 / \sqrt{2}) \, \transpose{(1 \ \ 1)}
\end{equation} 
respectively. With these two, the geometric phase obtained from the retarder is given as
\begin{equation}
   \gamma = \arg \big\{ \adjoint{\vv{E}}_a \vv{E}_b \big\} = 0 \, .
\end{equation} 

From this we can see that all three methods --- the graphical approach of \figref{fig:retarder_phase}, the mathematical approach of \eqref{eq:tan2L}, and the Jones-vector approach --- agree on the result.

\section{The effect of polarization components on geometric phase, II}\label{sec:retarder2}

Figure~\ref{fig:retarder_phase}(b) shows a second example, this time of a 22.5\degree linear polarization state propagating through the same linear retarder as in \figref{fig:retarder_phase}(a). This time, because the amplitude splitting is no longer symmetric, we will have a geometric phase even when the element is a quarter-wave plate.
\begin{description}
   \item[Step 1] In the input wave, the H-wave component is smaller than in \figref{fig:retarder_phase}(a), and the V-wave component is not zero.
   \item[Step 2] We project the 22.5\degree polarization state onto the (D,A) eigenbasis for the retarder.
   \item[Step 3] Apply the retardance: $+\delta/2$ to the D-wave, $-\delta/2$ to the A-wave.
   \item[Step 4] Compose the D- and A-waves into a single elliptically-polarized wave with peak location given by $\gamma$.
\end{description}
For the 22.5\degree input case, we see that the final wavefront position shifts towards the component wave with the larger amplitude, and this location is in general shifted by $\gamma$ from the initial reference plane. This is a ``geometric'' phase because it does not depend on propagation distance $z$ or time $t$, but rather depends only on the relationships between the input SOP, the eigenbasis of the retarder, and the retardance.

As with Sec.~\ref{sec:retarder}, we can make the graphical approach above quantitative as follows. When we decompose the initial 22.5\degree polarized wave into the (D,A) basis, we obtain
\begin{equation}
   \vv{E}_1 = \cos (\pi/8) \vv{E}\subrm{D} \, , \quad \vv{E}_2 = -\sin (\pi/8) \vv{E}\subrm{A} \, .
\end{equation} 
Once we apply the retardance, these become
\begin{equation}
   \vv{E}_1 = \cos (\tfrac{\pi}{8}) e^{-i (\delta-\pi) / 2} \vv{E}\subrm{D} \, , \quad \vv{E}_2 = \sin (\tfrac{\pi}{8}) e^{-i (\pi-\delta) / 2} \vv{E}\subrm{A} \, .
\end{equation} 
The elliptical wave created by composing these two has amplitude $\bar{A}$ and phase shift $\bar{\gamma}$ given by
\begin{equation}
   \bar{A}^2 = \cos^2 (\pi/8) + [-\sin (\pi/8)]^2 = 1
\end{equation} 
and
\begin{align}
   \tan \bar{\gamma} 
      &= \frac{\cos^2 (\tfrac{\pi}{8}) \sin ([\pi-\delta] / 2) - \sin^2 (\tfrac{\pi}{8}) \sin ([\pi-\delta] / 2)}{\cos^2 (\tfrac{\pi}{8}) \cos ([\pi-\delta] / 2) + \sin^2 (\tfrac{\pi}{8}) \cos ([\pi-\delta] / 2)} \notag \\
      &= -\frac{\tan [(\delta-\pi) / 2]}{\sqrt{2}} \, .
\end{align} 
Therefore, for a quarter waveplate retarder ($\delta = \pi/2$), we find that the geometric phase is $\bar{\gamma} = -0.615$ radians.

We also also follow the Jones vector method, for which the input and output waves $\vv{E}\subrm{a}$ and $\vv{E}\subrm{b}$ are
\begin{equation}
   \vv{E}_a = \begin{pmatrix} \cos (\pi/8) \\ -\sin (\pi/8) \end{pmatrix} \, , \quad \vv{E}_b = \begin{pmatrix} \cos (\pi/8) e^{-i (\delta-\pi) / 2} \\ \sin (\pi/8) e^{+i (\delta-\pi) / 2} \end{pmatrix} \, .
\end{equation} 
For $\delta = \pi/2$, these give
\begin{equation}
   \gamma = \arg \big\{ \adjoint{\vv{E}}_a \vv{E}_b \big\} = \SI{-0.615}{radians} \, .
\end{equation} 

In this example, the input polarization state does not split in equal amplitudes between the two eigenstates of the retarder. As a result, we find that the geometric phase is no longer zero, but that the peak of the elliptical wave emerging from the retarder is shifted by \SI{-0.615}{radians} with respect to the input reference plane.

\section{Interferogram detection of geometric phase}\label{sec:interferogram}

\begin{figure}[h]
    \centering
    \includegraphics[width=0.9\linewidth]{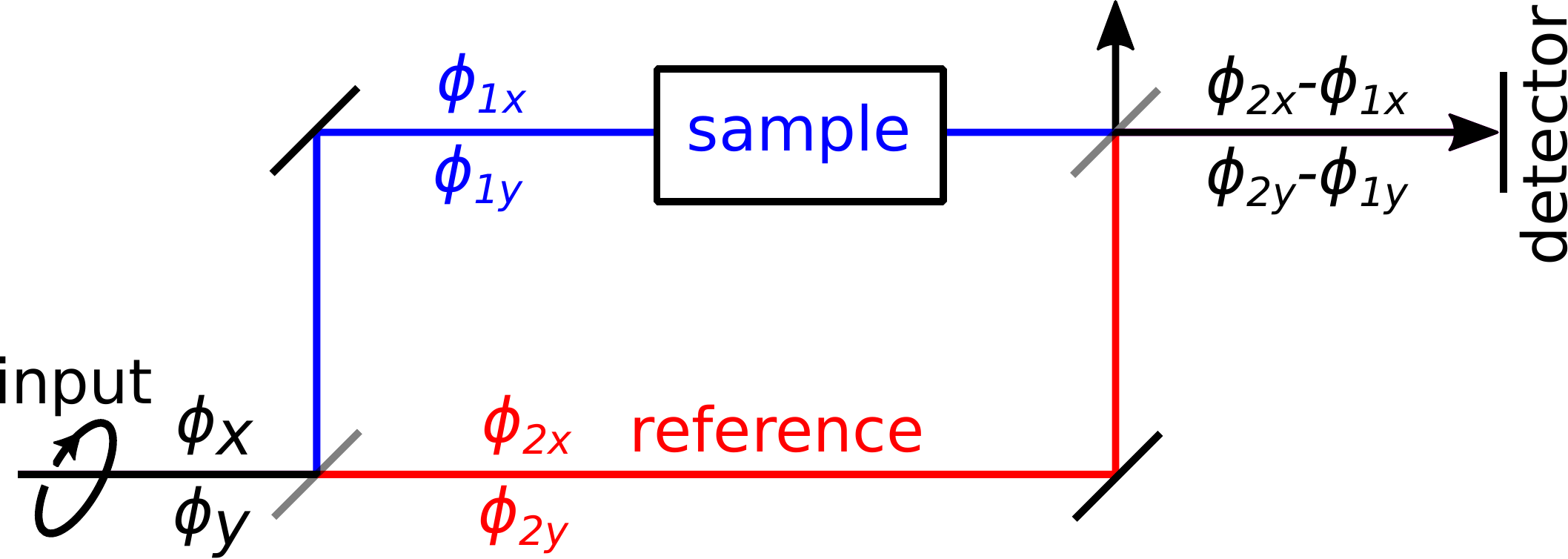}
    \figcaption{A Mach-Zender interferometer, showing the wave phases used in Eqs~\ref{eq:interferogram_waves}--\ref{eq:interferogram}.} 
    \label{fig:mach_zender}
\end{figure}

Measurement of optical geometric phase requires an interferometer, in order to detect the geometric phase as a shift with respect to a reference phase. This involves combining the output from the sample arm of the interferometer --- the one for which we are calculating the geometric phase --- with a wave transmitted through the reference arm of the interferometer. Figure~\ref{fig:mach_zender} shows the optical layout for a Mach-Zender interferometer --- the model we use below for considering interferometric measurement. In a typical setup, the polarization state of the reference arm of the interferometer is designed to match the SOP of the light output by the sample arm.

We write wave $\vv{E}_1$ as the electromagnetic wave from the sample arm, and $\vv{E}_2$ as the wave from the reference arm, using the Jones vectors
\begin{equation}\label{eq:interferogram_waves}
   \vv{E}_1 = e^{i \psi_1} \begin{pmatrix} A_{1x} e^{-i \phi_{1x}} \\ A_{1y} e^{+i \phi_{1y}} \end{pmatrix} , \quad \vv{E}_2 = e^{i \psi_2} \begin{pmatrix} A_{2x} e^{-i \phi_{2x}} \\ A_{2y} e^{+i \phi_{2y}} \end{pmatrix} \, ,
\end{equation}
where $\psi = kz - \omega t$ is the propagation phase (sometime called the ``dynamic phase'') obtained by the wave after passing through the interferometer, and $\phi$ encodes the phase delay between the $x$ and $y$ components of the wave. At this point, it is likely not at all clear that the sample arm encodes a geometric phase, but \ref{app:encoding} shows how the expression for $\vv{E}_1$ in \eqref{eq:interferogram_waves} can be expressed in a form to make the geometric phase $e^{i \gamma}$ explicit. The phase difference $\Delta = \psi_2 - \psi_1$ is the difference in optical path length between the two arms of the interferometer --- a degree of freedom that exists in the interferometer setup.

\vspace{\baselineskip}

When the sample and reference waves combine to produce the interferogram, the resulting intensity distribution is expressed by
\begin{align}
   I &= \adjoint{(\vv{E}_1 + \vv{E}_2)} \, (\vv{E}_1 + \vv{E}_2) \notag \\
      &= A_{1x}^2 + A_{2x}^2 + 2 A_{1x} A_{2x} \cos [(\psi_1 - \psi_2) - (\phi_{1x} - \phi_{2x})] \notag \\
      &\qquad + A_{1y}^2 + A_{2y}^2 + 2 A_{1y} A_{2y} \cos [(\psi_1 - \psi_2) - (\phi_{1y} - \phi_{2y})] \notag \\
      &= C + 2 A_{1x} A_{2x} \cos (\Delta - \delta_x) + 2 A_{1y} A_{2y} \cos (\Delta - \delta_y) \, , \label{eq:interferogram}
\end{align} 
where $C = A_{1x}^2 + A_{2x}^2 + A_{1y}^2 + A_{2y}^2$, $\delta_x = \phi_{1x} - \phi_{2x}$, and $\delta_y = \phi_{1y} - \phi_{2y}$. As a result of the adjoint operation, the interferogram $I$ contains none of the time dependence or $z$-dependence that the optical waves themselves have. The part that has disappeared from the expression is the mean dynamic phase of the two arms.

The interferogram consists of a constant $C$ plus two cosine waves that oscillate with respect to the variable $\Delta$. Thus, as we tune the OPD between the arms of the interferometer, the intensity modulates. Because we have a sum of two cosine waves, we can use the same approach that we did for the 1D wave composition formulas to calculate the phase of the interferogram peak with respect to $\Delta$. Taking the derivative of $I$ with respect to $\Delta$, setting the derivative to zero, and solving for $\Delta$, we obtain
\begin{equation}
   \tan \Delta = \frac{A_{1x} A_{2x} \sin \delta_x - A_{1y} A_{2y} \sin \delta_y}{A_{1x} A_{2x} \cos \delta_x + A_{1y} A_{2y} \cos \delta_y} \, . \label{eq:interferogram_phase}
\end{equation}
Although the result is no longer a propagating wave but rather a stationary interferogram, we find that the expression for the interferogram phase $\Delta$ is nearly the same as that of the geometric phase $\gamma$ of the 2D propagating wave, \eqref{eq:tan2L}. In fact, if $A_{1x} = A_{2x}$ and $A_{1y} = A_{2y}$, then the \eqref{eq:interferogram_phase} is exactly \eqref{eq:tan2L}. Thus, when the component amplitudes of the two arms are exactly matched, then the interferogram phase $\Delta$ exactly matches the propagating wave's geometric phase: $\tan \Delta = \tan \gamma$. Although it is not a requirement, this is easiest to see when the polarization states of the two arms are also exactly matched (i.e., $\delta_x = \delta_y$). In this special case, the interferogram expression \eqref{eq:interferogram} simplifies to
\begin{equation}
   I = C + C \cos \delta \, .
\end{equation} 

When the amplitudes in the two arms are not matched, and the polarization states are not the same, then the wave geometric phase and the interferogram phase are not the same. If the polarization states of the two arms are not the same, then attenuating one arm relative to the other will have an effect on the location of the peak. This is a feature that is not widely recognized in discussions about geometric phase, though it is a prominent element in Pancharatnam's original paper as a result of his focus on dichroic crystals while pursuing this work.\cite{Pancharatnam1956}

The results presented in \eqref{eq:interferogram_phase} allow us to describe what happens in the case of ``open loop'' configurations of geometric phase measurements. Such open loop situations --- where the polarization state output by the interferometer's sample arm is not matched to that in the reference arm --- are the subject of lengthy discussion in the literature, but we now have a straightforward means of quantifying what happens in this case, and how this affects the geometric phase measurement. The closed loop configuration, when attenuation in the sample arm is negligible, will produce a phase value that agrees with geometric phase predictions, but if attenuation is significantly different in the two arms, especially when the two polarization states differ, then the interferogram phase will diverge from $\gamma$.

The fact is that the interferogram is not a wave in the same way that light is a wave. The interferogram is stationary, has different units, and is not a transverse wave. Keeping this in mind, it is only natural to observe that they will not in general possess the same phase, except under specific measurement conditions.

Finally, we can add that if the geometric phase $\gamma$ is explicitly represented in the wave expression, such as by rewriting  \eqref{eq:interferogram_waves} in the form
\begin{equation}
   \vv{E}_1 = e^{i (\psi_1 + \gamma)} \begin{pmatrix} A_{1x} e^{-i \phi_{1x}} \\ A_{1y} e^{+i \phi_{1y}} \end{pmatrix} , \quad \vv{E}_2 = e^{i \psi_2} \begin{pmatrix} A_{2x} e^{-i \phi_{2x}} \\ A_{2y} e^{+i \phi_{2y}} \end{pmatrix} \, ,
\end{equation}
then we can see that the result \eqref{eq:interferogram_phase} merely has the geometric phase added to it,
\begin{equation}
   \tan (\Delta - \gamma) = \frac{A_{1x} A_{2x} \sin \delta_x - A_{1y} A_{2y} \sin \delta_y}{A_{1x} A_{2x} \cos \delta_x + A_{1y} A_{2y} \cos \delta_y} \, ,
\end{equation}
so that the geometric phase $\gamma$ from the sample arm produces one shift, while term on the right hand side of the equation represents an additional phase shift produced by the interference of two beams of differing polarization state.


\section{Conclusions}

Whereas previous work in the literature has discussed geometric phase in the abstract formalism of matrix calculus or spherical trigonometry, we have shown that the same results can be derived entirely from considering the phases of waves and analyzing how the wave peak position changes when waves are added together. This makes it possible to visualize phase relationships from the waves themselves, with minimal mathematical abstraction. 


Our approach provides an argument (Sec.~\ref{sec:2D}) for why the unusual quantities of ``instantaneous intensity''~\cite{Hannonen2019} (the square of the electric field, $E^2$) and ``instantaneous Stokes vector'' are used in the geometric phase literature. While previous researchers have used these without justification, other than that they work, we see that these arise naturally from calculating the position of the wave peak of a 2D polarized wave.

Whereas the existing literature calculates $\gamma$ only before and after a homogeneous optical element, it does not provide a model for calculating $\gamma$ continuously as a wave propagates through a polarization element. The wave model that we present here shows how this is to be constructed.

We also found that the likely reason why so much of the geometric phase literature assumes closed-loop cycles is that this is a situation for which the interferogram phase and the wave phase are identical. If the reference arm polarization state and the sample arm state are different, then the interferogram will exhibit a phase shift in accordance with \eqref{eq:interferogram_phase}.

Finally, our results also provide a simple explanation for the differences between geometric phases of scalar waves and vector waves~\cite{Aleksiejunas1997} --- differences entirely due to the disparity between composing two parallel waves, \eqref{eq:tan_phi3}, and composing two orthogonal waves, \eqref{eq:tan2L}.




\appendix

\section{Summing together $N$ waves in 1D}\label{app:Nwaves}

The Harmonic Addition theorem states that the sum of $N$ waves can itself be written as a wave~\cite{Hecht2002}
\begin{equation*}
   \psi = \sum_i A_i \cos (kz - \omega t + \phi_i) = \bar{A} \cos (kz - \omega t - \bar{\phi}) \, ,
\end{equation*}
with an amplitude and phase of
\begin{align*}
   \bar{A} &= \bigg[ \sum_{i=1}^N \sum_{j=1}^N A_i A_j \cos (\phi_i - \phi_j) \bigg]^{1/2} \\
      &= \bigg[ \sum_{i=1}^N A_i^2 + 2 \sum_{i=1}^N \sum_{j>i}^N A_i A_j \cos (\phi_i - \phi_j) \bigg]^{1/2} \, , \\
   \tan \bar{\phi} &= \frac{\sum_{i=1}^N A_i \sin \phi_i}{\sum_{i=1}^N A_i \cos \phi_i} \, .
\end{align*}

\section{Comparing Eq.~13 to an expression in the existing literature}\label{app:compare}

Equation~7 in Ref.~\citenum{Gutierrez-Vega2011}, and Equation~1 in Ref.~\citenum{Lopez-Mago2017} give
\begin{equation}\label{eq:Lopez-Mago-1}
   \Phi = \arg \big\{ \mu\subrm{p} + \mu\subrm{q} + (\mu\subrm{p} - \mu\subrm{q}) \vv{s}\subrm{p} \cdot \vv{s}\subrm{a} \big\} \, .
\end{equation} 
From the definitions given in those papers, we can write $\mu\subrm{p} = e^{-i \delta / 2}$ and $\mu\subrm{q} = e^{i \delta / 2}$. Also, we can note that $\vv{s} \cdot \vv{p} = \cos (2 \psi)$ where $\psi$ is an angle between the two points $\vv{s}\subrm{p}$ and $\vv{s}\subrm{a}$ on the Poincar{\'e} sphere. As a result, \eqref{eq:Lopez-Mago-1} becomes
\begin{align*}
   \Phi &= \arg \big\{ e^{-i \delta / 2} + e^{i \delta / 2} + (e^{-i \delta / 2} - e^{i \delta / 2}) \cos (2 \psi) \big\} \\
      &= \arctan \big[ \tan (\delta / 2) \cos (2 \psi) \big] \, .
\end{align*} 
In \eqref{eq:tan2L_symmetric}, we do not use the angle $\phi$ but rather use $A_1$ and $A_2$. However, if we represent these two amplitudes using an angle $\phi$, then $A_1 = \cos \phi$, $A_2 = \sin \phi$, and \eqref{eq:tan2L_symmetric} becomes
\begin{equation*}
   \tan \gamma = \tan (\delta) \, \frac{\cos^2 \phi - \sin^2 \phi}{\cos^2 \phi + \sin^2 \phi} 
      = \tan (\delta) \cos (2 \phi) \, .
\end{equation*}
Once we note that our definition for $\delta$ is based on the square of the electric field, whereas that of Ref.~\citenum{Lopez-Mago2017} is based on the Jones vectors, and thus the field amplitudes themselves, we can now see that the two expressions are exactly the same, though expressed in different notation.

\section{Hidden encoding of a global phase inside local phases}\label{app:encoding}

In order to see how a global phase can be encoded inside the apparently local phases defined in \eqref{eq:interferogram_waves}, we can consider the following example. In our Mach-Zender interferometer, we place a single retarder in the sample arm, with retardance $\delta$. The light incident on the sample arm is linearly polarized at 0\degree (i.e. an H-wave) and the retarder is also oriented with azimuth 0\degree. Then the input and output states in the sample arm will be
\begin{align}
   E_1 &= e^{i \psi_1} \begin{pmatrix} 1 \\ 0 \end{pmatrix} \, , \\
   E_2 &= e^{i \psi_2} \begin{pmatrix} 1 \cdot e^{i \delta / 2} \\ 0 \cdot e^{-i \delta / 2} \end{pmatrix} = e^{i [\psi_2 + (\delta / 2)]} \begin{pmatrix} 1 \\ 0 \end{pmatrix} \, .
\end{align} 
Thus, the polarization state is unchanged, but we have delayed the H-wave by half the retardance.

Next, we consider the case when the input polarization state is oriented at 45\degree (i.e., a D-wave), and the retarder fast-axis is also oriented with an azimuth angle of 45\degree. Thus, this setup is effectively equivalent as the previous case, but with our reference axis rotated. Still working in the $x$-$y$ basis, the input and output waves of this case will be
\begin{equation}
   E_1 = e^{i \psi_1} \begin{pmatrix} \cos 45\degree \\ \sin 45\degree \end{pmatrix} \, , \qquad E_2 = e^{i \psi_2} \begin{pmatrix} \cos 45\degree e^{i \delta / 2} \\ \sin 45\degree e^{-i \delta / 2} \end{pmatrix} \, .
\end{equation} 
The retardance $\delta$ in the output wave no longer takes on the appearance of a global phase, but is instead encoded within the local phases of the $x$ and $y$ components.




\end{document}